\documentclass[intlimits,twoside,a4paper]{article}
\usepackage{amsmath,amssymb}
\usepackage{graphicx}
\usepackage{wrapfig}
\usepackage[T2A]{fontenc}
\usepackage[cp1251]{inputenc}

\usepackage{cmpj2}

\issue{2012}{15}{4}{43003}

\doinumber{10.5488/CMP.15.43003}

\title[Spin-1/2 Ising-Heisenberg model in the external magnetic field]
{Magnetization process in the exactly solved
spin-1/2 Ising-Heisenberg model on  decorated Bethe lattices}

\author[J. Stre\v{c}ka, C. Ekiz]{J. Stre\v{c}ka\refaddr{label1}, \ C. Ekiz\refaddr{label2}}
\addresses{
\addr{label1} Department of Theoretical Physics and Astrophysics, Faculty of Science,
P.J. \v{S}af\'{a}rik University, \\ Park Angelinum 9, 040 01 Ko\v{s}ice, Slovak Republic
\addr{label2} Department of Physics, Faculty of Science, Adnan Menderes University,
090 10 Ayd\i n, Turkey}

\date{Received June 22, 2012, in final form August 10, 2012}
\authorcopyright{J. Stre\v{c}ka, C. Ekiz, 2012}

\begin{document}

\maketitle

\begin{abstract}
The spin-1/2 Ising-Heisenberg model on diamond-like decorated Bethe lattices is exactly solved in the presence of the longitudinal magnetic field by combining the decoration-iteration mapping transformation with the method of exact recursion relations. In particular, the ground state and low-temperature magnetization process of the ferrimagnetic version of the considered model is investigated in detail. Three different magnetization scenarios with up to two consecutive fractional magnetization plateaus
were found, whereas the intermediate magnetization plateau may either correspond to the classical ferrimagnetic spin arrangement and/or the field-induced quantum ferrimagnetic spin ordering without any classical counterpart.

\pacs 05.50.+q, 75.10.-b, 75.10.Jm, 75.10.Kt 75.40.Cx, 75.60.Ej
\end{abstract}

\section{Introduction}

Low-dimensional quantum spin systems have attracted much attention over the past few decades, since they exhibit a lot of striking quantum phenomena including fractional magnetization plateaus, spin-Peierls dimerization, unconventional spin-liquid ground states, or many other peculiar valence-bond-solid ground states such as the Haldane phase~\cite{mil01,lac11}. It is worth noting that the most remarkable experimental findings reported for low-dimensional spin systems were mostly satisfactorily interpreted with the help of quantum Heisenberg model and its various extensions. From the theoretical point of view, an exact treatment of the quantum Heisenberg model remains an unresolved problem mainly due to substantial mathematical difficulties, which arise from a noncommutability of spin operators involved in the relevant Hamiltonian. However, this mathematical complexity can be avoided by considering simpler Ising-Heisenberg models, which describe hybrid classical-quantum spin systems constituted both by the `classical' Ising as well as the quantum Heisenberg spins. The hybrid Ising-Heisenberg models can be exactly treated by making use of generalized mapping transformations,
which were originally introduced by Syozi~\cite{syo51,syo72} and later on generalized by Fisher~\cite{fis59}, Rojas et al.~\cite{roj09,roj11} and
one of the present authors~\cite{str10}.

In this work, the generalized decoration-iteration transformation is combined with the method of exact recursion relations in order to obtain exact results for the spin-$\frac{1}{2}$ Ising-Heisenberg model on diamond-like decorated Bethe lattices in the presence of the longitudinal magnetic field. It should be noted that the applied decoration-iteration transformation establishes a rigorous mapping equivalence between the investigated model system and the spin-$\frac{1}{2}$ Ising model on a corresponding simple Bethe lattice with the effective nearest-neighbour interaction and the effective magnetic field. Owing to this precise mapping correspondence, exact results for the spin-$\frac{1}{2}$ Ising-Heisenberg model on the diamond-like decorated Bethe lattices can be subsequently extracted from the relevant exact solution of the spin-$\frac{1}{2}$ Ising model on a simple Bethe lattice by means of the method of exact recursion relations~\cite{bax82,tho82,muk74}.

The organization of this paper is as follows. In section~\ref{model}, the detailed description of the investigated model system is presented together with the basic steps
of its exact solution. The most interesting results are then presented and discussed in section~\ref{result}. In particular, our attention is focused on the ground state  and low-temperature magnetization process of the ferrimagnetic version of the considered model. Finally, some concluding remarks are drawn in section~\ref{conclusion}.

\section{Ising-Heisenberg model on decorated Bethe lattices}
\label{model}

\begin{figure}[!ht]
\begin{center}
\includegraphics[width=12cm]{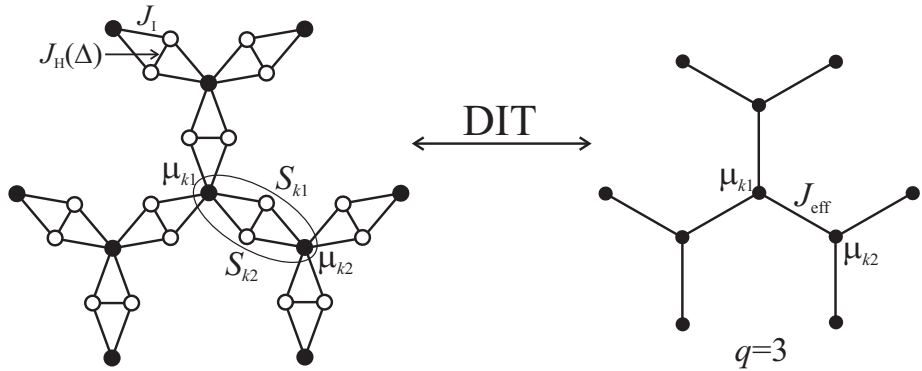}
\end{center}
\vspace{-0.6cm}
\caption{The spin-$\frac{1}{2}$ Ising-Heisenberg model on the diamond-like decorated Bethe lattice (figure on the left) and its exact mapping via the decoration-iteration transformation (DIT) onto the spin-$\frac{1}{2}$ Ising model on a simple Bethe lattice (figure on the right). The full (empty) circles denote lattice positions of the Ising (Heisenberg) spins, the ellipse demarcates the elementary diamond-shaped spin cluster described by the $k$th bond Hamiltonian (\ref{eq:2}).}
\label{fig1}
\end{figure}

Let us introduce the spin-$\frac{1}{2}$ Ising-Heisenberg model on a diamond-like decorated Bethe lattice, which is schematically illustrated on the left-hand-side of figure~\ref{fig1} on the particular example of the underlying Bethe lattice with the coordination number $q=3$. In this figure, the full circles label lattice positions of the Ising spins $\mu=\frac{1}{2}$, while the empty circles mark lattice positions of the Heisenberg spins $S=\frac{1}{2}$. One may infer from figure~\ref{fig1} that the magnetic structure of the investigated model is formed by the Ising spins placed at lattice sites of a deep interior of infinite Cayley tree (Bethe lattice), which are linked together through the Heisenberg spin pairs placed in-between each couple of the Ising spins. The total Hamiltonian of the spin-$\frac{1}{2}$ Ising-Heisenberg model on diamond-like decorated Bethe lattices reads
\begin{eqnarray}
{\cal H} = - J_{\rm H} \sum_{(k,l)}^{Nq/2} \left[\Delta \left(S_{k}^{x} S_{l}^{x} + S_{k}^{y} S_{l}^{y} \right) + S_{k}^{z} S_{l}^{z} \right]
           - J_{\rm I} \sum_{(k,i)}^{2Nq} S_{k}^{z} \mu_{i}^{z} - H_{\rm A} \sum_{i=1}^{N} \mu _{i}^{z} - H_{\rm B} \sum_{k=1}^{Nq} S_{k}^{z}\,.
\label{eq:1}
\end{eqnarray}
Here, $S_{k}^{\alpha} \,\, (\alpha = x,y,z)$ and $\mu^{z}_{i}$ represent spatial components of the spin-$\frac{1}{2}$ operator, the parameter $J_{\rm H}$ denotes the XXZ interaction between the nearest-neighbour Heisenberg spins, the parameter $\Delta$ controls a spatial anisotropy in this interaction between the easy-axis $(\Delta < 1)$
and easy-plane $(\Delta > 1)$ regime, and the parameter $J_{\rm I}$ marks the Ising interaction between the nearest-neighbour Heisenberg and Ising spins, respectively. Furthermore, two Zeeman's terms $H_{\rm A}$ and $H_{\rm B}$ determine the magnetostatic energy of the Ising and Heisenberg spins in a longitudinal magnetic field.

It is quite evident from figure~\ref{fig1} that each pair of Heisenberg spins is surrounded by one couple of the Ising spins located at lattice sites of a simple Bethe lattice and hence, the model under consideration can alternatively be viewed as the Bethe lattice of Ising spins whose (fictitious) bonds are decorated in a diamond-like fashion by two quantum Heisenberg spins. In view of further manipulations, it is, therefore, of practical importance to rewrite the total Hamiltonian~(\ref{eq:1}) as a sum of bond Hamiltonians
\begin{eqnarray}
{\cal H}=\sum_{k=1}^{Nq/2} {\cal H}_{k}\,,
\label{eq:1b}
\end{eqnarray}
whereas the bond Hamiltonian ${\cal H}_{k}$ involves all the interaction terms belonging to the $k$th diamond-shaped cluster specifically delimited in figure~\ref{fig1} by an ellipse
\begin{eqnarray}
{\cal H}_{k} &=  &-  J_{\rm H} \left[\Delta \left(S_{k1}^x S_{k2}^x + S_{k1}^y S_{k2}^y \right) + S_{k1}^z S_{k2}^z \right]
                           - J_{\rm I} \left( S_{k1}^z + S_{k2}^z \right) \left(\mu_{k1}^z +\mu_{k2}^z \right)  \nonumber\\
    &&-  H_{\rm B} \left(S_{k1}^{z}+S_{k2}^{z} \right) - \frac{H_{\rm A}}{q} \left(\mu _{k1}^{z} + \mu _{k2}^{z} \right).
\label{eq:2}
\end{eqnarray}
Owing to the validity of the commutation relationship between different bond Hamiltonians $[{\cal H}_{i}, {\cal H}_{j}] = 0$, the partition function
can be partially factorized into a product of bond partition functions
\begin{equation}
{\cal Z}_{\rm IHM} = \sum_{\{ \mu_i \}} \prod_{k = 1}^{Nq/2} \mbox{Tr}_k \exp(- \beta {\cal H}_k)
               = \sum_{\{ \mu_i \}} \prod_{k = 1}^{Nq/2} {\cal Z}_{k}\,,
\label{eq:3}
\end{equation}
where $\beta=1/(k_{\rm B}T)$, $k_{\rm B}$ is the Boltzmann's constant and $T$ is the absolute temperature. The symbol $\mbox{Tr}_{k}$ denotes a trace over degrees of freedom of two Heisenberg spins from the $k$th diamond-shaped cluster and the summation $\sum_{\{\mu _{i}\}}$ runs over all possible configurations of the Ising spins. The bond partition function ${\cal Z}_{k}$ can be evaluated in the most straightforward way by a direct diagonalization of the bond Hamiltonian~(\ref{eq:2}) within the particular subspace of the  $k$th Heisenberg spin pair and employing a trace invariance of the bond partition function with respect to a unitary transformation. After executing this procedure one gains
the resultant expression, which implies a possibility of applying the generalized decoration-iteration transformation~\cite{fis59,roj09,roj11,str10}
\begin{eqnarray}
{\cal Z}_k  \left(\mu_{k1}^z, \mu_{k2}^z \right) &=& 2 \exp \left[\frac{\beta H_{\rm A}}{q} \left(\mu _{k1}^{z} + \mu _{k2}^{z} \right) \right]
\left\{ \exp \left( \frac{\beta J_{\rm H}}{4} \right) \cosh \left[\beta J_{\rm I} \left(\mu_{k1}^z + \mu_{k2}^z \right) + \beta H_{\rm B} \right] \right.   \nonumber\\
&& \left. + \exp \left(-\frac{\beta J_{\rm H}}{4} \right) \cosh \left(\frac{\beta J_{\rm H}\Delta}{2} \right) \right \}=
A \exp \left[\beta J_{\rm eff} \mu_{k1}^z \mu_{k2}^z + \frac{\beta H_{\rm eff}}{q} \left(\mu_{k1}^z + \mu_{k2}^z \right) \right].
\label{eq:4}
\end{eqnarray}
Considering four available combinations of spin states of two Ising spins $\mu_{k1}^z$ and $\mu_{k2}^z$, one gets from the transformation formula~(\ref{eq:4}) three independent equations that unambiguously determine the mapping parameters $A$, $J_{\rm eff}$ and $H_{\rm eff}$
\begin{eqnarray}
A = 2 \left(V_{+}V_{-}V_{0}^2 \right)^{1/4}, \qquad
\beta J_{\rm eff} = \ln \left(\frac{V_{+}V_{-}}{V_{0}^2} \right), \qquad
\beta H_{\rm eff} = \beta H_{A} + \frac{q}{2} \ln \left(\frac{V_{+}}{V_{-}} \right),
\label{eq:5}
\end{eqnarray}
which are for simplicity defined through the functions $V_{\pm}$ and $V_{0}$
\begin{eqnarray}
V_{\pm} \!\!\!&=&\!\!\!
          \exp \left( \frac{\beta J_{\rm H}}{4} \right) \cosh \left(\beta J_{\rm I} \pm \beta H_{\rm B} \right)
        + \exp \left(-\frac{\beta J_{\rm H}}{4} \right) \cosh \left(\frac{\beta J_{\rm H} \Delta}{2} \right),\nonumber \\
V_{0} \!\!\!&=&\!\!\!
          \exp \left( \frac{\beta J_{\rm H}}{4} \right) \cosh \left(\beta H_{\rm  B} \right)
        + \exp \left(-\frac{\beta J_{\rm H}}{4} \right) \cosh \left(\frac{\beta J_{\rm H} \Delta}{2} \right).
\label{eq:6}
\end{eqnarray}
At this stage, the direct substitution of the algebraic mapping transformation~(\ref{eq:4}) into the factorized form
of the partition function~(\ref{eq:3}) leads to a rigorous mapping relationship
\begin{equation}
{\cal Z}_{\rm IHM} (\beta, J_{\rm I}, J_{\rm H}, \Delta, H_{\rm A}, H_{\rm B}, q) = A^{\frac{Nq}{2}} {\cal Z}_{\rm IM} (\beta, J_{\rm eff}, H_{\rm eff}, q),
\label{eq:7}
\end{equation}
which connects the partition function ${\cal Z}_{\rm IHM}$ of the spin-$\frac{1}{2}$ Ising-Heisenberg model on the diamond-like decorated Bethe lattice
with the partition function ${\cal Z}_{\rm IM}$ of the spin-$\frac{1}{2}$ Ising model on a corresponding simple (undecorated) Bethe lattice
schematically illustrated on the right-hand-side of figure~\ref{fig1} and mathematically given by the Hamiltonian
\begin{equation}
{\cal H}_{\rm IM} = - J_{\rm eff} \sum_{(i,j)}^{Nq/2} \mu_{i}^{z} \mu_{j}^{z} - H_{\rm eff} \sum_{i=1}^{N} \mu _{i}^{z}\,.
\label{hami}
\end{equation}
Apparently, the mapping parameters $J_{\rm eff}$ and $H_{\rm eff}$ given by equations (\ref{eq:5})--(\ref{eq:6}) determine the effective nearest-neighbour interaction and the effective magnetic field of the corresponding spin-$\frac{1}{2}$ Ising model on the simple Bethe lattice, while the mapping parameter $A$ is just a simple multiplicative factor
in the established mapping relation~(\ref{eq:7}) between both partition functions.

Now, other physical quantities of our particular interest follow quite straightforwardly. For instance,  with the help of equation~(\ref{eq:7}), one easily finds a similar mapping relation between the free energy $F_{\rm IHM}$ of the spin-$\frac{1}{2}$ Ising-Heisenberg model on the diamond-like decorated Bethe lattice and the free energy
$F_{\rm IM}$ of the equivalent spin-$\frac{1}{2}$ Ising model on a simple Bethe lattice
\begin{equation}
F_{\rm IHM} = - k_{\rm B} T \ln {\cal Z}_{\rm IHM} = F_{\rm IM} - \frac{N q k_{\rm B} T}{2} \ln A.
\label{eq:11}
\end{equation}
Consequently, the single-site sublattice magnetization of the Ising spins can be calculated by differentiating the free energy~(\ref{eq:11})
with respect to the relevant magnetic field $H_{\rm A}$
\begin{eqnarray}
m_{\rm A} = - \frac{1}{N} \frac{\partial F_{\rm IHM}}{\partial H_{\rm A}}
= - \frac{1}{N} \left( \frac{\partial F_{\rm IM}}{\partial \beta H_{\rm eff}} \right)
\frac{\partial \beta H_{\rm eff}}{\partial H_{\rm A}} = m_{\rm IM} (\beta, J_{\rm eff}, H_{\rm eff}).
\label{eq:8}
\end{eqnarray}
According to equation~(\ref{eq:8}), the sublattice magnetization of the Ising spins in the spin-$\frac{1}{2}$ Ising-Heisenberg model on the diamond-like decorated Bethe lattice is equal to the magnetization of the corresponding spin-$\frac{1}{2}$ Ising model on the simple Bethe lattice with the effective nearest-neighbour interaction $J_{\rm eff}$ and the effective magnetic field $H_{\rm eff}$ given by (\ref{eq:5})--(\ref{eq:6}). A similar calculation procedure can also be performed for obtaining the single-site sublattice magnetization of the Heisenberg spins, which can be for convenience expressed in terms of the magnetization $m_{\rm IM}$ and the nearest-neighbour pair correlation  $\varepsilon_{\rm IM}$ of the equivalent spin-$\frac{1}{2}$ Ising model on the simple Bethe lattice
\begin{eqnarray}
m_{\rm B}&=&
- \frac{1}{Nq} \frac{\partial F_{\rm IHM}}{\partial H_{\rm B}}  = \frac{1}{2} \frac{\partial \ln A}{\partial \beta H_{\rm B}}
- \left( \frac{1}{Nq} \frac{\partial F_{\rm IM}}{\partial \beta J_{\rm eff}} \right) \frac{\partial \beta J_{\rm eff}}{\partial H_{\rm B}}
- \left( \frac{1}{Nq} \frac{\partial F_{\rm IM}}{\partial \beta H_{\rm eff}} \right) \frac{\partial \beta H_{\rm eff}}{\partial H_{\rm B}}  \nonumber\\
&=& \frac{1}{8} \left (\frac{W_{+}}{V_{+}} - \frac{W_{-}}{V_{-}} + 2\frac{W_{0}}{V_{0}} \right)
+ \frac{\varepsilon_{\rm IM}}{2} \left (\frac{W_{+}}{V_{+}} - \frac{W_{-}}{V_{-}} - 2\frac{W_{0}}{V_{0}} \right)
+ \frac{m_{\rm IM}}{2} \left(\frac{W_{+}}{V_{+}} + \frac{W_{-}}{V_{-}} \right).
\label{eq:15}
\end{eqnarray}
The newly defined functions $W_{\pm}$ and $W_0$ are given by
\begin{eqnarray}
W_{\pm} = \exp \left( \frac{\beta J_{\rm H}}{4} \right) \sinh \left(\beta J_{\rm I} \pm \beta H_{\rm B} \right), \qquad
W_{0} = \exp \left( \frac{\beta J_{\rm H}}{4} \right) \sinh \left(\beta H_{\rm B} \right).
\label{w}
\end{eqnarray}

To complete our exact calculation of both sublattice magnetizations, it is now sufficient to substitute into the derived formulas (\ref{eq:8})--(\ref{eq:15}) the relevant exact results for the magnetization and nearest-neighbour spin-spin correlation of the corresponding spin-$\frac{1}{2}$ Ising model on the simple Bethe lattice with the effective nearest-neighbour interaction $J_{\rm eff}$ and the effective magnetic field $H_{\rm eff}$ given by (\ref{eq:5})--(\ref{eq:6}). The sublattice magnetization and spin-spin correlation function of the spin-1/2 Ising model on the undecorated Bethe lattice can be rigorously found within the framework of exact recursion relations~\cite{bax82,tho82,muk74,oha07,izm98}. If the simple Bethe lattice (see figure~\ref{fig1}, figure on the right) is `cut' at a central site with the spin $\mu _{k1}$, it will disintegrate into $q$ identical branches and the partition function of the system will take the form
\begin{eqnarray}
{\cal Z} = \sum_{\mu_{k1}} \exp(\beta H_{\rm eff}\mu_{k1}) \left[ g_{n}(\mu_{k1} ) \right ]^{q},
\label{eq:16}
\end{eqnarray}
where $g_{n}(\mu_{k1})$ is the partition function of a separate branch
\begin{eqnarray}
g_{n}(\mu_{k1})=\sum_{\mu_{k2} } \exp (\beta J_{\rm eff} \mu_{k1}\mu_{k2}+\beta H_{\rm eff}\mu_{k2})\left [ g_{n-1}(\mu_{k2}) \right ]^{q-1}.
\label{eq:17}
\end{eqnarray}
By using of (\ref{eq:17}), one can easily obtain a recursion relationship for the variable $x_{n}=\frac{g_{n}(-1/2)}{g_{n}(+1/2)}$
\begin{equation}
x_{n}  =  \frac{\exp \left(-\frac{\beta J_{\rm eff}}{4} + \frac{\beta H_{\rm eff}}{2}\right) + \exp\left(\frac{\beta J_{\rm eff}}{4} - \frac{\beta H_{\rm eff}}{2}\right) x_{n-1}^{q-1}}
               {\exp\left(\frac{\beta J_{\rm eff}}{4} + \frac{\beta H_{\rm eff}}{2}\right) + \exp\left(-\frac{\beta J_{\rm eff}}{4} - \frac{\beta H_{\rm eff}}{2}\right) x_{n-1}^{q-1}}\,.
\label{eq:10}
\end{equation}
Even though the parameter $x_n$ does not have a direct physical sense, it plays a crucial role in determining the canonical ensemble averages of all physical quantities
in the limit $n \to \infty$. For instance, one easily obtains the following expressions for the magnetization and nearest-neighbour spin-spin correlation function of the spin-1/2 Ising model on the Bethe lattice
\begin{eqnarray}
m_{\rm IM} \!\!\!&=&\!\!\! \frac{1}{2} \frac{\exp \left(\beta H_{\rm eff} \right) -  x^q}{\exp \left(\beta H_{\rm eff} \right) + x^q} \,,  \nonumber\\
\varepsilon_{\rm IM} \!\!\!&=&\!\!\! \frac{1}{4}
\frac{\exp \left(\frac{\beta J_{\rm eff}}{4} + \beta H_{\rm eff} \right) - 2 \exp \left(-\frac{\beta J_{\rm eff}}{4} \right) x^{q-1}
    + \exp \left(\frac{\beta J_{\rm eff}}{4} - \beta H_{\rm eff} \right) x^{2q-2}}
     {\exp \left(\frac{\beta J_{\rm eff}}{4} + \beta H_{\rm eff} \right) + 2 \exp \left(-\frac{\beta J_{\rm eff}}{4} \right) x^{q-1}
    + \exp \left(\frac{\beta J_{\rm eff}}{4} - \beta H_{\rm eff} \right) x^{2q-2}}\, .
\label{eq:9}
\end{eqnarray}
which can be both expressed through a stable fixed point $x = \displaystyle \lim_{n \to \infty} x_n$ of the recurrence relation~(\ref{eq:10}).

\section{Results and discussion}
\label{result}

\begin{wrapfigure}{t}{0.5\textwidth}
\vspace*{-0.5cm}
\centerline{\includegraphics[width=0.5\textwidth]{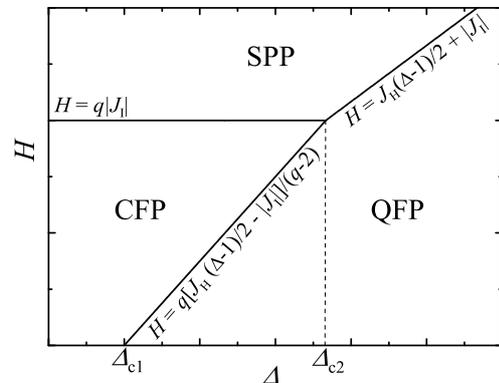}}
\vspace*{-0.8cm}
\caption{The general ground-state phase diagram in the $\Delta-H$ plane.}
\vspace*{-0.2cm}
\label{fig2}
\end{wrapfigure}

In this part, let us proceed to a discussion of the most interesting results obtained for the ferrimagnetic version of the spin-$\frac{1}{2}$ Ising-Heisenberg model
on the diamond-like decorated Bethe lattice with the ferromagnetic Heisenberg interaction $J_{\rm H} > 0$ and the antiferromagnetic Ising interaction $J_{\rm I} < 0$,
which,  at sufficiently low fields, will favour the antiparallel alignment between the nearest-neighbouring Ising and Heisenberg spins, respectively.
It is worthwhile to remark that the critical behaviour of the considered model in the absence of the external magnetic field has been investigated in some detail
in our previous work~\cite{eki10} and hence, the effect of a non-zero magnetic field will be at the main focus of our research interest. To reduce
the total number of free parameters, the most notable features of the magnetization process will be illustrated for a specific choice $H \equiv H_{\rm A} = H_{\rm B}$,
which coincides with setting Land\'e g-factors of the Ising and Heisenberg spins equal to each other.

First, let us comment on possible spin arrangements emerging at zero temperature. Owing to the validity of the commutation relationship between the different cluster Hamiltonians,
the ground-state spin arrangements can easily be obtained by searching for the lowest-energy eigenstate of the cluster Hamiltonian~(\ref{eq:2}). The ground-state phase diagram displayed in figure~\ref{fig2} implies the existence of three different ground states, which can be thoroughly characterized by the following eigenvectors
\begin{eqnarray}
| \mbox{CFP} \rangle \!\!\!&=&\!\!\! \prod_{k=1}^N \left| \mu_{k}^z = - \frac{1}{2} \right\rangle \prod_{k=1}^{Nq/2} \left| S_{k1}^z = \frac{1}{2} \right\rangle
\left| S_{k2}^z = \frac{1}{2} \right\rangle,
 \nonumber\\
| \mbox{QFP} \rangle \!\!\!&=&\!\!\! \prod_{k=1}^N \left| \mu_{k}^z = \mbox{sgn}(H) \frac{1}{2} \right\rangle \prod_{k=1}^{Nq/2} \frac{1}{\sqrt{2}} \left(
  \left| S_{k1}^z = \frac{1}{2} \right \rangle \left| S_{k2}^z = -\frac{1}{2} \right\rangle
+ \left| S_{k1}^z = -\frac{1}{2} \right\rangle \left| S_{k2}^z = \frac{1}{2}  \right\rangle \right),
\nonumber \\
| \mbox{SPP} \rangle \!\!\!&=&\!\!\! \prod_{k=1}^N \left| \mu_{k}^z = \frac{1}{2} \right\rangle \prod_{k=1}^{Nq/2} \left| S_{k1}^z = \frac{1}{2} \right\rangle
\left| S_{k2}^z = \frac{1}{2} \right\rangle.
\label{gs3}
\end{eqnarray}
As could be expected, two ground states correspond to classical spin arrangements with a perfect parallel and antiparallel alignments between the nearest-neighbour Ising and Heisenberg spins to be further referred to as the classical ferrimagnetic phase (CFP) and the saturated paramagnetic phase (SPP), respectively. Apart from those
rather trivial phases, one may also detect a more spectacular quantum frustrated phase (QFP) with a peculiar spin frustration of the Ising spins stemming
from a quantum entanglement of the Heisenberg spin pairs. As a matter of fact, the emergent quantum superposition of two possible antiferromagnetic states of the Heisenberg
spin pairs is responsible in QFP for a complete randomness of the Ising spins at a zero magnetic field as convincingly evidenced in our previous study~\cite{eki10}.
Due to the spin frustration, all the Ising spins tend to align into the external-field direction for arbitrary but non-zero magnetic field and, consequently,
a striking \textit{quantum ferrimagnetic phase} develops from QFP with a full polarization of the Ising spins and the non-magnetic nature of the Heisenberg spin pairs.
The existence of QFP alone seems to be a quite general feature of the Ising-Heisenberg models, where a mutual competition between the easy-axis Ising interaction and the easy-plane Heisenberg interaction takes place~\cite{str06,eki11}. Furthermore, all phase transitions between three available ground states are of the first order and their explicit form is given in figure~\ref{fig2} along the depicted phase boundaries.

\begin{figure}[ht]
\centerline{\includegraphics[width=0.45\textwidth]{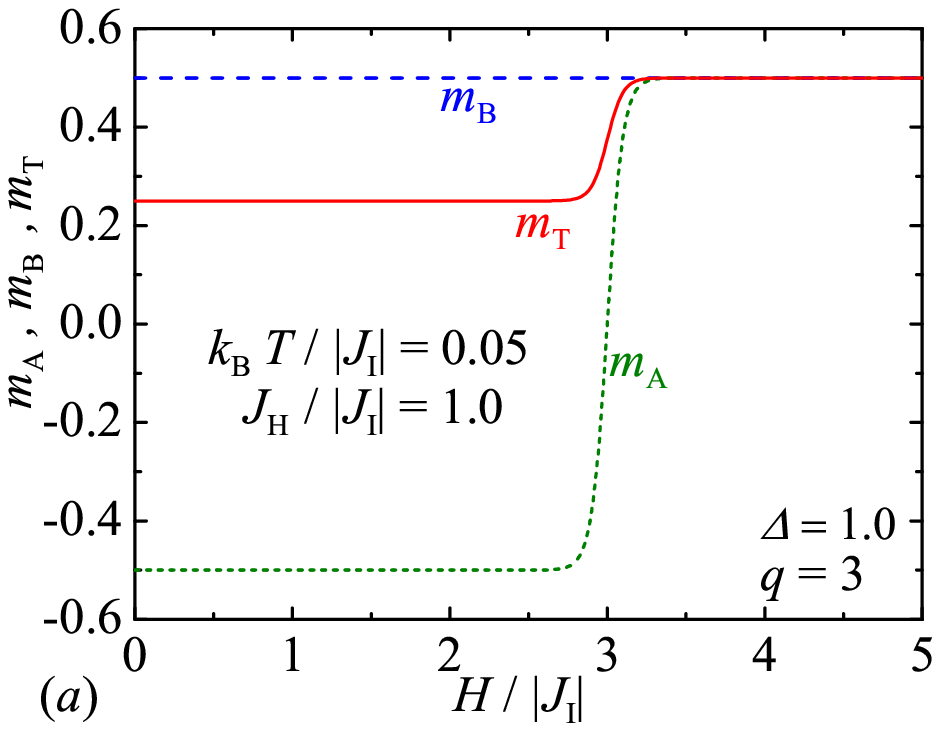}
            \includegraphics[width=0.45\textwidth]{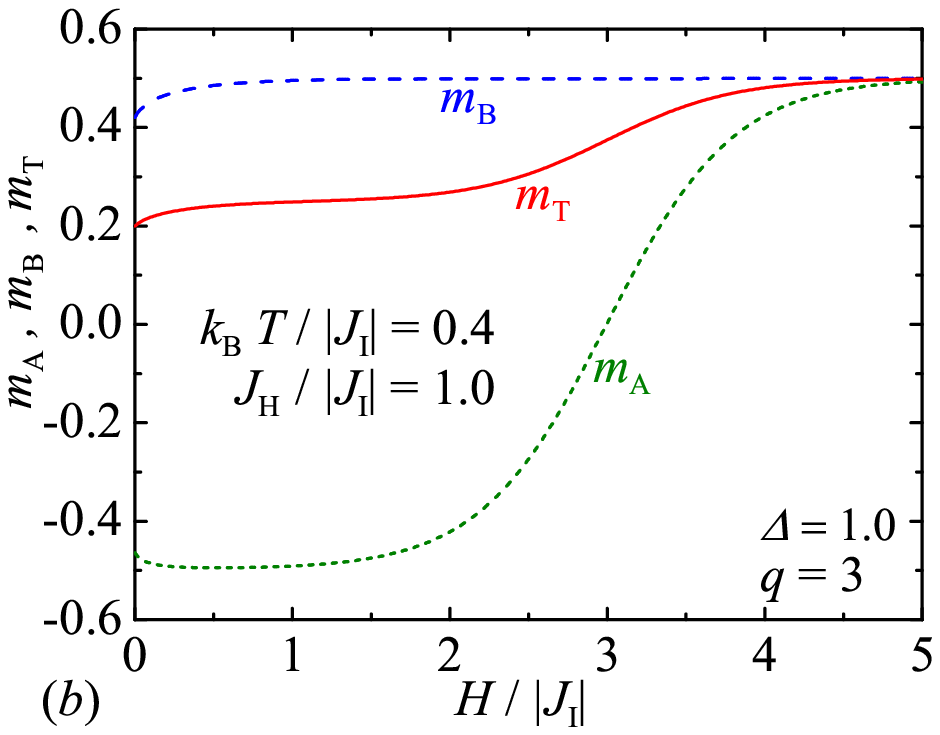}}
            \vspace*{-0.4cm}
\centerline{\includegraphics[width=0.45\textwidth]{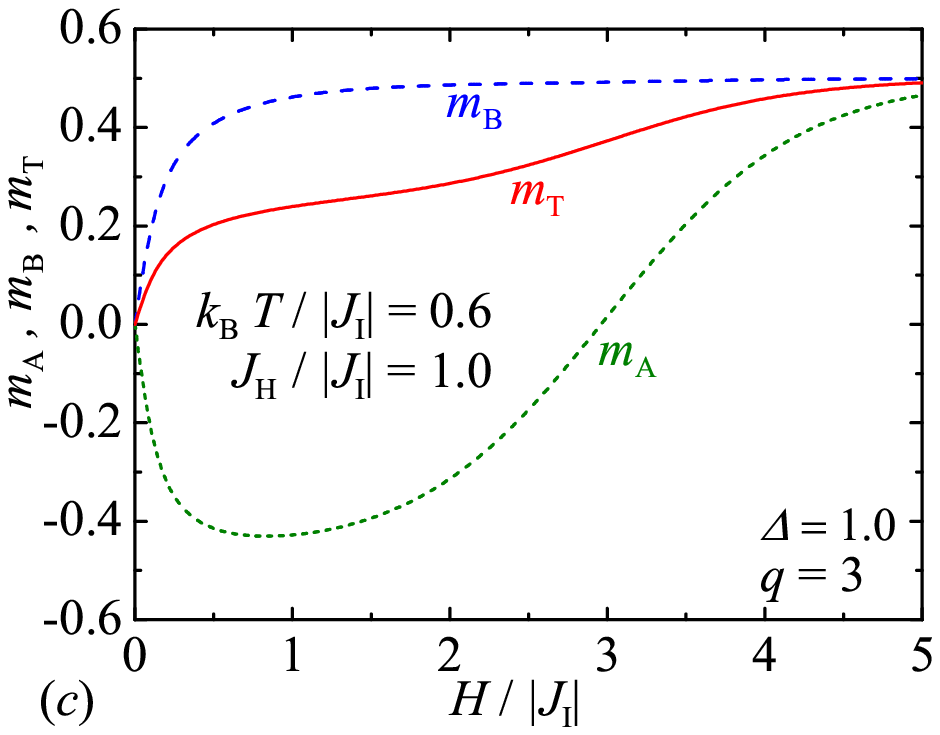}
            \includegraphics[width=0.45\textwidth]{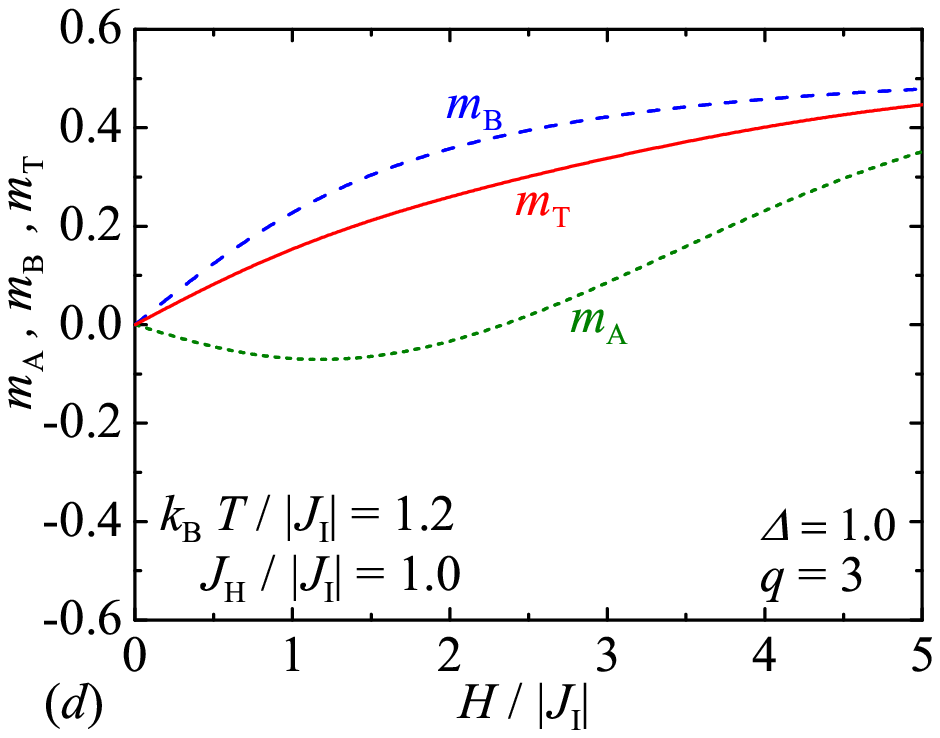}}
\vspace*{-0.5cm}
\caption{(Color online) The total and sublattice magnetizations as a function of the external magnetic field for the spin-$\frac{1}{2}$ Ising-Heisenberg model
with the coordination number $q=3$, the interaction ratio $J_{\rm H}/|J_{\rm I}| = 1.0$, the exchange anisotropy $\Delta=1.0$ and four different temperatures.}
\label{fig3}
\end{figure}
\begin{figure}[!t]
\centerline{\includegraphics[width=0.45\textwidth]{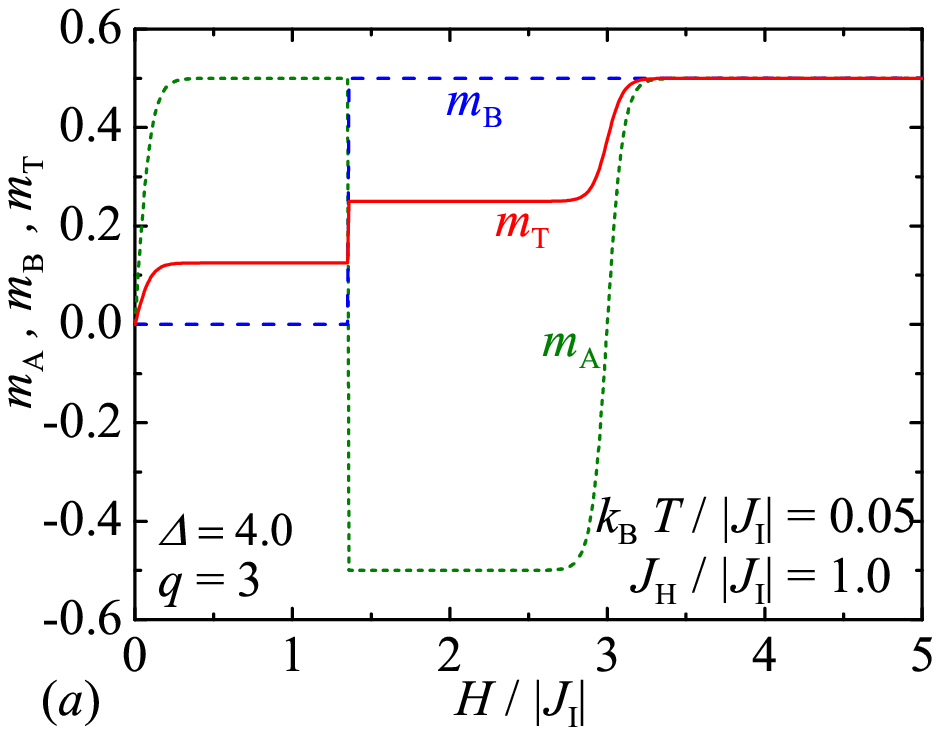}
            \includegraphics[width=0.45\textwidth]{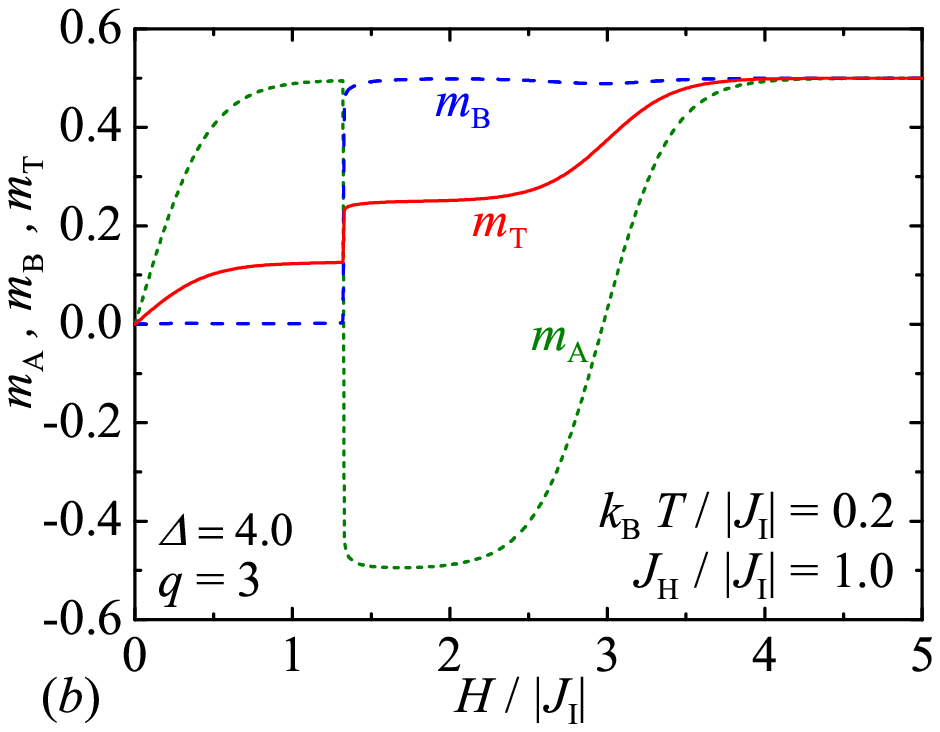}}
            \vspace*{-0.4cm}
\centerline{\includegraphics[width=0.45\textwidth]{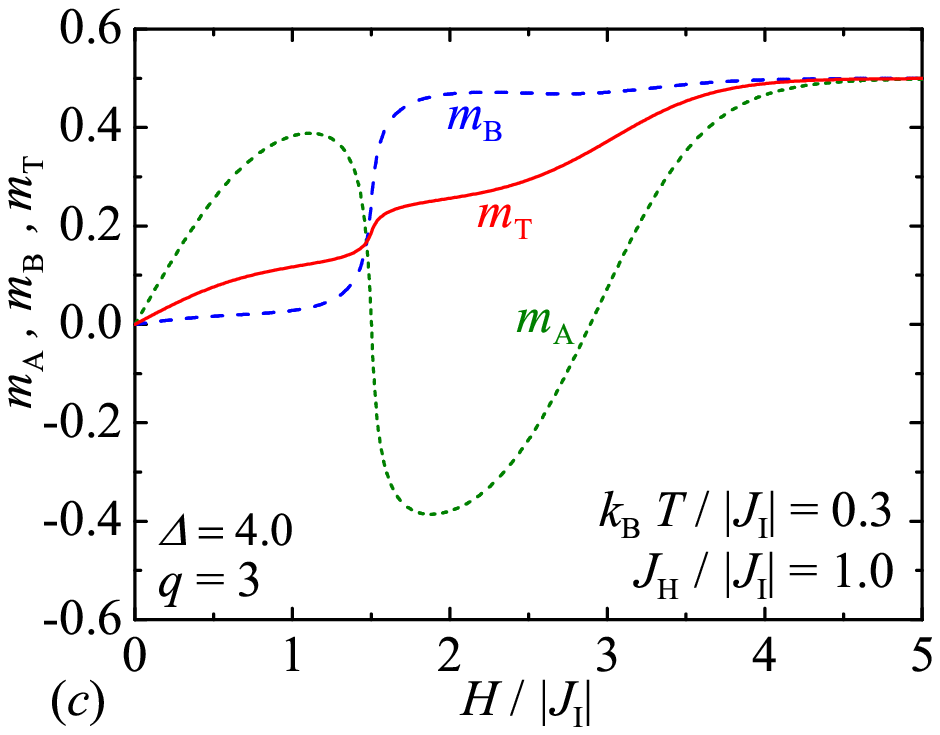}
            \includegraphics[width=0.45\textwidth]{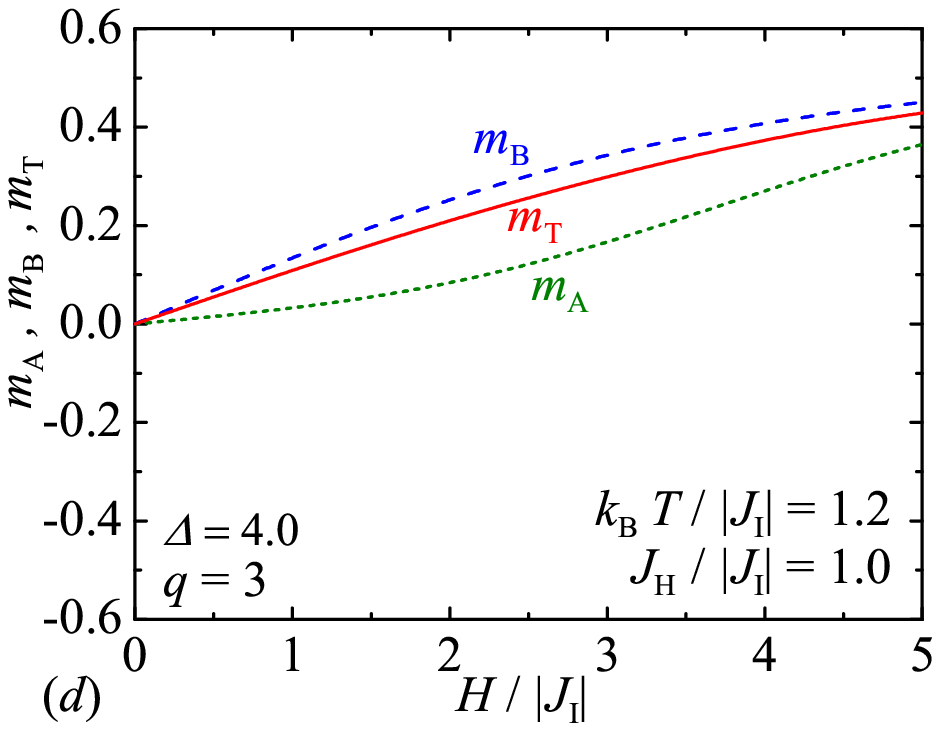}}
\vspace*{-0.5cm}
\caption{(Color online) The total and sublattice magnetizations as a function of the external magnetic field for the spin-$\frac{1}{2}$ Ising-Heisenberg model
with the coordination number $q=3$, the interaction ratio $J_{\rm H}/|J_{\rm I}| = 1.0$, the exchange anisotropy $\Delta=4.0$ and four different temperatures.}
\label{fig4}
\end{figure}
\begin{figure}[!b]
\vspace{-1cm}
\centerline{\includegraphics[width=0.45\textwidth]{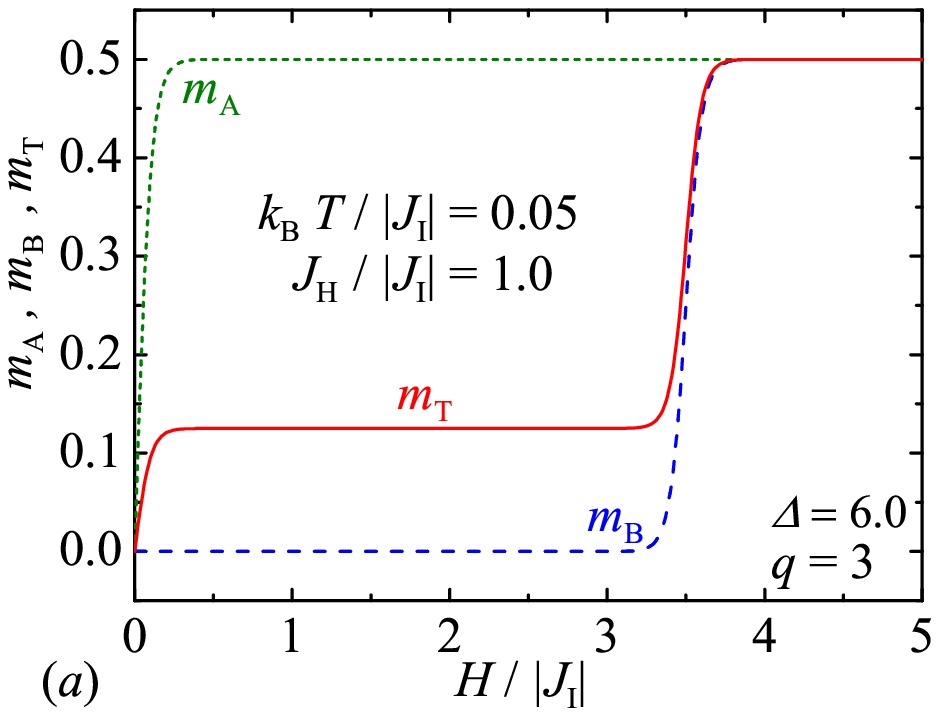}
            \includegraphics[width=0.45\textwidth]{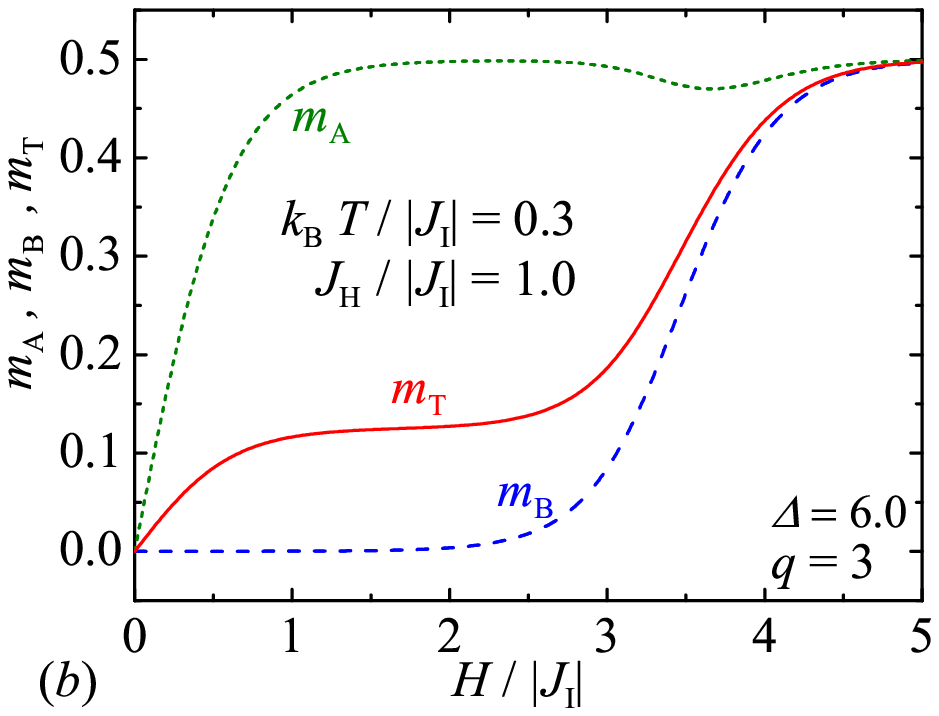}}
            \vspace*{-0.4cm}
\centerline{\includegraphics[width=0.45\textwidth]{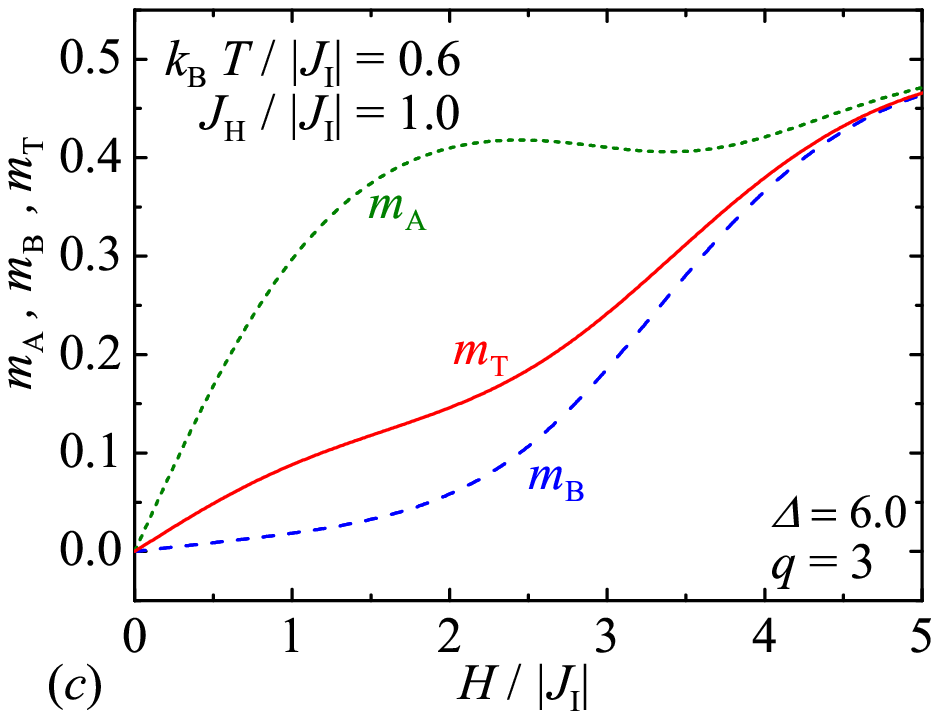}
            \includegraphics[width=0.45\textwidth]{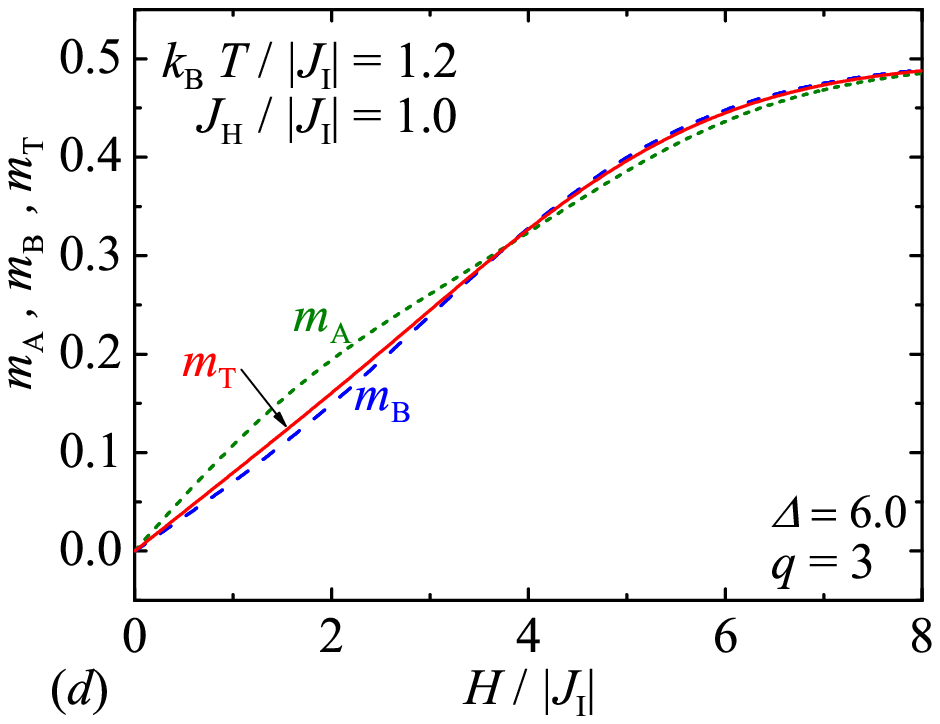}}
\vspace*{-0.5cm}
\caption{(Color online) The total and sublattice magnetizations as a function of the external magnetic field for the spin-$\frac{1}{2}$ Ising-Heisenberg model
with the coordination number $q=3$, the interaction ratio $J_{\rm H}/|J_{\rm I}| = 1.0$, the exchange anisotropy $\Delta=6.0$ and four different temperatures.}
\label{fig5}
\end{figure}

Now, let us illustrate typical magnetization scenarios as displayed in figures~\ref{fig3}--\ref{fig5} for the spin-$\frac{1}{2}$ Ising-Heisenberg model on the diamond-like decorated Bethe lattice with the coordination number $q=3$, the specific value of the interaction ratio $J_{\rm H}/|J_{\rm I}| = 1.0$, three different values of the exchange anisotropy $\Delta$ and several temperatures. It is worthwhile to remark that the total single-site magnetization $m_{\rm T} \equiv \left(m_{\rm A} + q m_{\rm B}\right)/(1 + q)$ is also plotted in figures~\ref{fig3}--\ref{fig5} in addition to both sublattice magnetizations $m_{\rm A}$ and  $m_{\rm B}$ of the Ising and Heisenberg spins, respectively. If the exchange anisotropy is selected below its first critical value $\Delta < \Delta_{\rm c1} \equiv 1 + 2 |J_{\rm I}|/J_{\rm H}$, then, one encounters a rather typical magnetization curve reflecting the field-induced transition from CFP to SPP as shown in figure~\ref{fig3}. It is quite clear that the  intermediate magnetization plateau observed at a half of the saturation magnetization indeed corresponds to the classical ferrimagnetic spin arrangement inherent to CFP and the magnetization plateau gradually diminishes upon increasing the temperature. The most significant changes in the displayed magnetization curve evidently occur if the temperature is selected slightly above the critical temperature of CFP (note that $k_{\rm B} T_{\rm c}/|J_{\rm I}| \approx 0.5$ for $\Delta=1$). Even though both sublattice magnetizations already start from zero in this particular case, they obviously tend towards typical magnetization values for CFP still bearing evidence of an intermediate magnetization plateau at moderate fields and temperatures [see figure~\ref{fig3}~(c)].


However, the most interesting magnetization process can be found if the exchange anisotropy is selected from the interval $\Delta \in (\Delta_{\rm c1}, \Delta_{\rm c2})$
with $\Delta_{\rm c2} \equiv 1 + 2 (q-1) |J_{\rm I}|/J_{\rm H}$. Under this condition, at low enough temperatures, the total magnetization exhibits two successive fractional
magnetization plateaus at one quarter and one half of the saturation magnetization [see figures~\ref{fig4}~(a)--(b)], which end up at two different field-induced transitions from QFP to CFP and, respectively, from CFP to SPP. The lower magnetization plateau at one quarter of the saturation magnetization gives a clear evidence of QFP, because the total magnetization starts from zero and it becomes non-zero mainly due to the field-induced alignment of the frustrated Ising spins. Moreover, it is quite interesting to observe
from figure~\ref{fig4} that the former field-induced transition between QFP and CFP is much sharper at a given temperature than the latter field-induced transition between CFP and SPP. Of course, the relevant magnetization curve becomes smoother upon increasing temperature until both magnetization plateaus completely disappear from the magnetization process above a certain temperature ($k_{\rm B} T/|J_{\rm I}| \approx 0.5$ for $\Delta = 4.0$).

Last but not least, the magnetization curve without the higher intermediate magnetization plateau at a half of the saturation magnetization can be detected whenever the exchange anisotropy exceeds its second critical value $\Delta > \Delta_{\rm c2}$. In agreement with the ground-state phase diagram shown in figure~\ref{fig2}, the low-temperature magnetization curve displays a direct field-induced transition from QFP towards SPP without passing through another magnetization plateau CFP. For illustration, the magnetization scenario of this type is depicted in figure~\ref{fig5}. It is worth noting that the field-induced polarization of the Heisenberg spins, which appears in the vicinity of the saturation field, may cause, at moderate temperatures, a transient lowering of the sublattice magnetization of the Ising spins as it can be clearly seen in figures~\ref{fig5}~(b)--(c). The partial lowering of the sublattice magnetization of the Ising spins can be attributed to a spin reorientation of the Heisenberg spins towards the external-field direction and a tendency of the nearest-neighbour Ising and Heisenberg spins to align antiparallel with respect to each other due to the antiferromagnetic interaction in-between them. Furthermore,
figure~\ref{fig5}~(d) shows an interesting crossing of both sublattice magnetizations, which occurs at sufficiently high temperatures on account of the antiferromagnetic correlations between the nearest-neighbour Ising and Heisenberg spins.

\begin{figure}[ht]
\centerline{\includegraphics[width=0.3\textwidth]{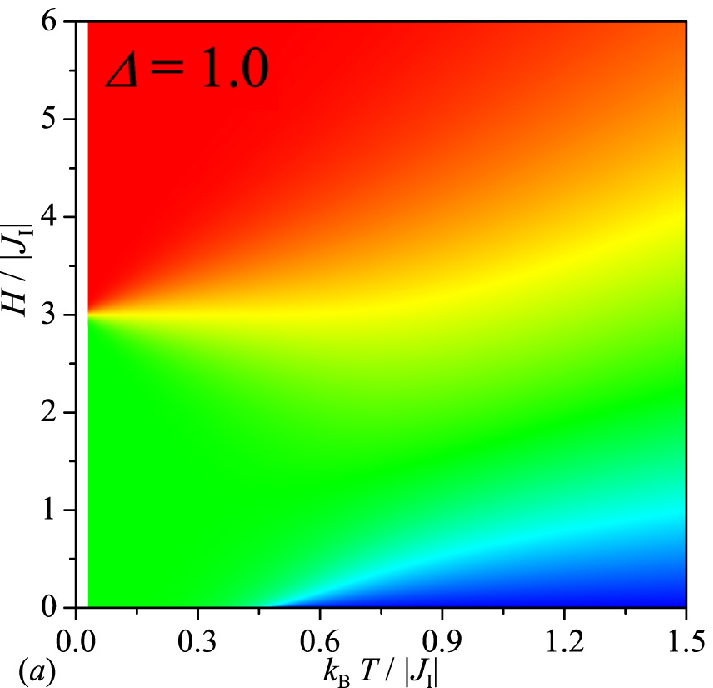}
            \includegraphics[width=0.3\textwidth]{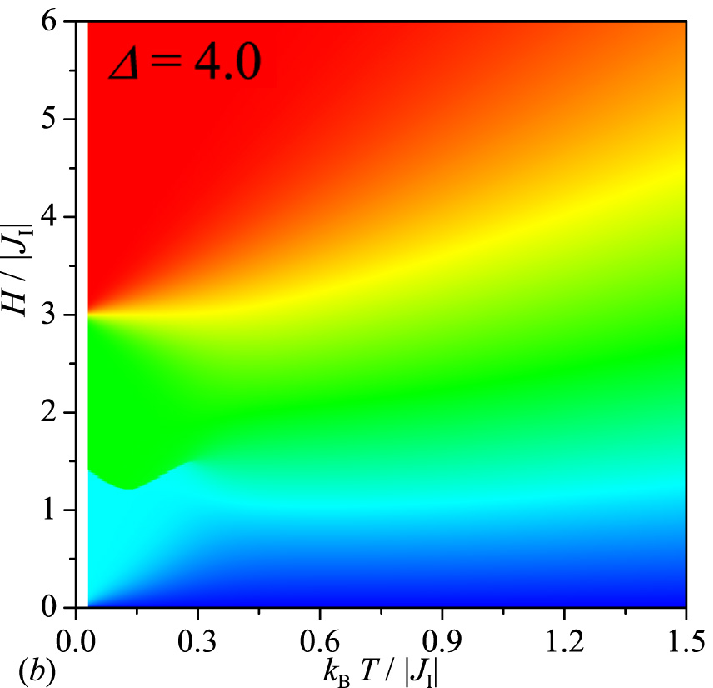}
            \includegraphics[width=0.36\textwidth]{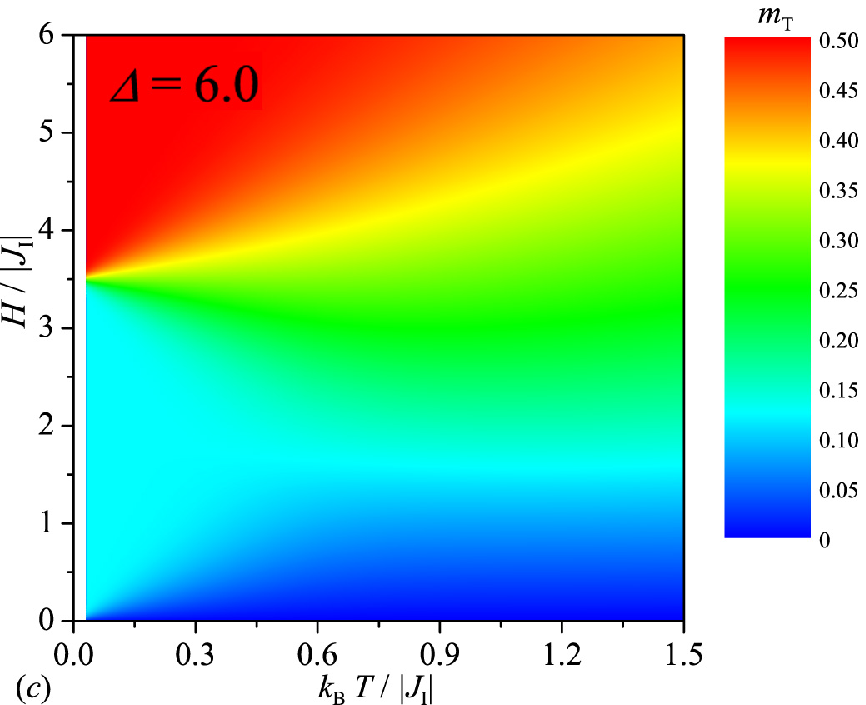}}
\caption{(Color online) A color map of the total magnetization as a function of the dimensionless temperature and external magnetic field for the spin-$\frac{1}{2}$ Ising-Heisenberg model on the diamond-like decorated Bethe lattice with the coordination number $q=3$, the interaction ratio $J_{\rm H}/|J_{\rm I}| = 1.0$ and three different values of the exchange anisotropy: (a) $\Delta=1.0$; (b) $\Delta=4.0$; (c) $\Delta=6.0$.}
\label{fig6}
\end{figure}

Let us conclude our analysis of the magnetization process by few comments on a color map of the total magnetization depicted in figure~\ref{fig6} as a function of temperature and external magnetic field. According to a unique color map labelling used in figure~\ref{fig6}, two fractional values of the total magnetization $m_{\rm T} = 0.125$ and $0.25$ that correspond to the intermediate magnetization plateaus associated with the appearance of QFP and CFP are displayed by cyan and green color, respectively. As could be expected, the quite extensive green region in figure~\ref{fig6}~(a) indicates a rather wide magnetization plateau at a half of the saturation magnetization emerging for relatively weak exchange anisotropies  $\Delta < \Delta_{\rm c1}$, while the wide cyan region in figure~\ref{fig6}~(c) implies the existence of a relatively robust magnetization plateau at one quarter of the saturation magnetization for strong enough exchange anisotropies $\Delta > \Delta_{\rm c2}$. Hence, if follows that the most striking magnetization profile  with two successive intermediate magnetization plateaus might be indeed expected for the intermediate exchange anisotropies $\Delta \in (\Delta_{\rm c1}, \Delta_{\rm c2})$. In fact, figure~\ref{fig6}~(b) serves in evidence of the presence of both intermediate magnetization plateaus, which are gradually smudged by thermal fluctuations as temperature increases. Interestingly, it turns out that the lower fractional magnetization plateau pertinent to QFP diminishes much more steadily with an increasing temperature in comparison  with the higher fractional magnetization plateau pertinent to CFP, which seems to be much more resistant against thermal fluctuations.

\section{Conclusion}
\label{conclusion}

The present work deals with the spin-$\frac{1}{2}$ Ising-Heisenberg model on diamond-like decorated Bethe lattices in the presence of the longitudinal magnetic field.
Exact solution for the investigated model has been obtained by combining the decoration-iteration mapping transformation with the method of exact recursion relations.
The former transformation method makes it possible to establish a rigorous mapping relationship with the equivalent spin-$\frac{1}{2}$ Ising model on a simple Bethe lattice,
which is subsequently exactly treated within the framework of the latter method based on exact recursion relations. Exact results for the partition function,
Gibbs free energy, total and both sublattice magnetizations were derived by making use of this rigorous approach.

Our particular attention was focused on exploring the ground state and low-temperature magnetization process of the ferrimagnetic version of the  model considered.
The most interesting finding stemming from our present study is an exact evidence of a rather diverse magnetization process. As a matter of fact, we have demonstrated
three different magnetization scenarios with up to two different fractional magnetization plateaus, whereas the intermediate magnetization plateau may either correspond
to the classical ferrimagnetic spin arrangement and/or the quantum ferrimagnetic spin ordering without any classical counterpart. The origin of the striking quantum ferrimagnetic phase lies in a peculiar spin frustration of the Ising spins, which comes from the nonmagnetic nature of the Heisenberg spin pairs governed by the symmetric quantum superposition of their two intrinsically antiferromagnetic spin states.

\section*{Acknowledgements}
This work was supported by the Scientific Grant Agency of Ministry of Education of Slovak Republic
under the VEGA Grant No.~1/0234/12 and by ERDF EU (European Union European regional development fund) grant
under the contract ITMS 26220120005 (activity 3.2.).

\newpage

\ukrainianpart

\title{Процес намагніченості в точно розв'язній спін-1/2 моделі Ізинга-Гайзенберга на декорованих гратках Бете}

\author{Й. Стречка\refaddr{label1}, \ С. Екіз\refaddr{label2}}

\addresses{
\addr{label1} Природничий факультет, Унiверситет iм. П.Й. Шафарика, Кошiце, Словацька республiка
\addr{label2} Природничий факультет, Університет ім. Аднана Мендереса, Айдин 090 10, Туреччина}

\makeukrtitle

\begin{abstract}
\tolerance=3000%
Спін-1/2 модель Ізинга-Гайзенберга на ромбоподібній декорованій гратці Бете розв'язано точно у присутності поздовжнього магнітного поля, поєднуючи декораційно-ітераційне перетворення з методом точних рекурсивних співвідношень. Зокрема, детально досліджено основний стан і низькотемпературне намагнічення феримагнітної версії розглянутої моделі. Знайдено три різні сценарії намагніченості з щонайбільше двома послідовними дробовими плато, де проміжне плато намагніченості може відповідати класичному феримагнітному спіновому впорядкуванню та/або індукованому полем квантовому феримагнітному спіновому впорядкуванню, яке не має жодного класичного аналога.
\keywords модель Ізинга-Гайзенберга, гратка Бете, точні результати, плато намагніченості

\end{abstract}

\end{document}